
\input phyzzx
\PHYSREV
\hbox to 6.5truein{\hfill UPR-0558T}
\hbox to 6.5truein{\hfill March, 1993}
 \title{PHENOMENOLOGICAL DESCRIPTION OF AN UNSTABLE FERMION}
\author{Jiang Liu}
\address{Department of Physics, University of Pennsylvania, Philadelphia,
         PA 19104}
$$ $$
\abstract
We present a phenomenological description for an unstable fermion
based upon one-loop renormalization of quantum field theory.
It is emphasized that wave function renormalization can introduce
important $CP$-conserving  and $CP$-violating phases.
Implications for the study of $CP$ violation are examined.
Applications are
given to $CP$-violating asymmetries in the  $t$ decays: $t\to bW^+,
bH^+$, in which we show that a naive calculation following the standard
rule  either is incomplete or violates $CPT$.

PACS\#: 11.30.Er, 13.20.Jf, 13.30.Ce.
\endpage

\chapter{Introduction}

An important
effect introduced by the instability of an unstable particle is to
shift the pole location of its propagator  to a complex value
$$m\to m-{i\over 2}\Gamma,\eqno(1.1)$$
where $m$ and $\Gamma$ are respectively the mass and  width of the particle.
In this paper we emphasize that
particle instability
also introduces important complications into the dynamics of particle
wave functions.

Consider the $CP$-violating asymmetries in the $t$ decays
$$\eqalign{
&\Delta_W\equiv \Gamma(t\to bW^+)-\Gamma(\bar t\to \bar bW^-),\cr
&\Delta_H\equiv \Gamma(t\to bH^+)-\Gamma(\bar t\to \bar bH^-),\cr}\eqno(1.2)$$
in a model\Ref\Lee{T. D. Lee, Phys. Rev. D8, 1226 (1973);
         S. Weinberg, Phys. Rev. Lett. 37, 657 (1976);
         J. Liu and L. Wolfenstein, Nucl. Phys. B289, 1 (1987).\splitout
         For a more complete reference see: {\it Higgs Hunter's
         Guide }, J. F. Gunion, H. E. Haber, G. Kane and S. Dawson,
         (Addison-Wesley) 1990.}
containing a light charged-Higgs-boson $H^{\pm}$
$(M_H<m_t-m_b$) with an interaction
$${\cal{L}}_I={g\over\sqrt 2}W^-_{\mu}\bar b\gamma^{\mu}Lt+
       {g\over \sqrt 2M_W}H^-\bar b(xm_bL+ym_tR)t+h.c.,\eqno(1.3)$$
where $L(R)={1\over 2}(1\mp\gamma_5)$ and $x$ and $y$ are dimensionless
parameters.  A relative phase of $x$ and $y$ violates $CP$.
For simplicity we ignore  family mixings.

The $CP$ asymmetries are generated from
the interference of the tree-level amplitudes with higher order corrections
to the vertex,
mass and wave function
$$\Delta_{W,H}=\Delta_{W,H}(vertex)+\Delta_{W,H}(mass)+\Delta_{W,H}(
    wave\ function).\eqno(1.4)$$
To lowest order of $\alpha=e^2/4\pi$, $t$ has only two
decay modes with final states
$bW^+$ and $bH^+$.   They are related
by final-state interactions.
Adjoining these  on-shell final-state
interactions to $t\to bW^+, bH^+$ corresponds to a
calculation of an absorptive part of a vertex correction.
The interference of the vertex corrections with
the tree-level amplitudes produces
a $CP$ asymmetry with
$$\Delta_W(vertex)=-\Delta_H(vertex).\eqno(1.5)$$
This relation
follows
 because only the `off-diagonal' final-state interaction
$bW^+\leftrightarrow bH^+$
(see Figs. 1c and 1d) is relevant\rlap.\Ref\Wolfen{
           L. Wolfenstein, Phys. Rev. D43, 1 (1991).  It is straightforward to
           generalize  the result given in this paper
           to situations in which final-state interactions also
           violate $CP$. }

One may attempt to calculate the
mass-renormalization contributions
by applying a rule\Ref\NP{
           M. Nowakowski and A. Pilaftsis, preprint, MZ-TH/92-16 (1992).}
suggested by (1.1) to the interaction lagrangian
$$\eqalign{& m_t\to m_t-{i\over 2}\Gamma_t,\cr
           & M_W\to M_W-{i\over 2}\Gamma_W.\cr}\eqno(1.6)$$
While the $CP$-conserving phase $i\Gamma_W$ in $M_W$ factors and hence
does not contribute, replacing $m_t$ in (1.3) by
$m_t-i\Gamma_t/2$ yields to lowest order in $x$ and $y$
$$\eqalign{&\Delta_W(mass)=0,\cr
&\Delta_H(mass)={g^2m_b^2Im(x^*y)\Gamma(t\to bW)\over
      16\pi M_W^2m_t^2}\sqrt{\lambda(m_t^2, M_H^2, m_b^2)},\cr
}\eqno(1.7)$$
where $\lambda(u,v,z)=u^2+v^2+z^2-2uv-2uz-2vz$.

Up to an overall phase arising from a chiral rotation,
the standard
wave-function renormalization respects $CP$ from $CPT$ invariance.
We will return to this point later.
It then follows  that if we were to use the standard Dirac algebra
to compute
the decay  rate, we would find
$$\Delta_W(wave\  function)=\Delta_H(wave\  function)=0,\eqno(1.8)$$
indicating (together with Eqs. (1.5) and (1.7)) that
$\Gamma(t\to all)-\Gamma(\bar t\to all)=\Delta_H(mass)\ne 0$,
in violation of $CPT$\rlap.\Ref\GHS{
          For discussions of $CPT$ constraints on $CP$-violating
          asymmetries in other processes see:
          J.-M. Gerard and W.-S. Hou, Phys. Rev. Lett. 62, 855 (1989),
          Phys. Rev. D43, 2909 (1991);  J. M. Soares, Phys. Rev. Lett.
          68, 2102 (1992);
          J. Liu, Phys. Rev. D47, R1741 (1993).}

Besides the application of (1.6) being questionable,
this seemingly trivial calculation does raise two important questions
relevant in general to the study of heavy fermion decays.
First, in view of (1.1) it is not clear whether  the standard
Dirac algebra
such as
$$\sum_{\beta=\pm}u_{\beta}(P)\otimes \bar u_{\beta}(P)=
{P\!\!\!\!/+m\over 2m}.\eqno(1.9)$$
still holds for a renormalized unstable spinor which has
an absorptive part of order $\alpha$.
Second, the conventional interpretation that
$\bar t$ in the $h.c.$ part of ${\cal{L}}_I$ is the
hermitian adjoint of $t$ may well be ambiguous
after radiative corrections, since  the one-loop
renormalized
lagrangian may not be  hermitian
because of particle instability. Obviously,
corrections   associated with these complications are
purely imaginary and of order $\Gamma_t\sim {\cal{O}}(\alpha)$
in amplitudes.
In terms of physical observables their effects (if any)
are at most of order $\alpha^2$ in the absence of $CP$
violation\rlap.\Ref\JGLIU{ In some cases, an order
$\alpha$ effect can show up in terms of certain spin-momentum correlations,
see for example:  J. Liu, UPR-0517T (1992) and references therein.}
However,
if $CP$ is violated,
the interference of $CP$-conserving
and $CP$-violating phases
may produce a $CP$-violating effect which is of order $\alpha$.
The purpose of this paper
is to investigate how this order $\alpha$ effect can
be treated consistently.

The rest part of this paper is organized as follows.
In the next section,
we study the lowest order dynamical behavior of an unstable fermion
by examining renormalization effects on its propagator.
We develop a phenomenological description for an unstable fermion
in section 3.  Applications are given to the $t$ decays to show
how $CPT$ invariance can be restored.
Also, we demonstrate how to compute
contributions arising from wave function renormalization.
Conclusions are presented in section 4.  Some technical
details are summarized in two appendices.

\chapter{Dynamics of An Unstable Fermion}

The lowest order dynamical behavior of a particle is
determined by its propagator.
Consider the $t$-quark as an example. The
tree level  result is
$$iS_{(0)}(P)={i\over P\!\!\!\!/-m_t^{(0)}}.\eqno(2.1)$$
The one-loop self-energy can be parameterized as
$$-\Sigma(P)=2A(P)P\!\!\!\!/L+2B(P)P\!\!\!\!/R
            +m_tC(P)L+D(P)m_tR,\eqno(2.2)$$
where $A, B, C$ and $D$ are form factors specified by
a given theoretical model.
$CPT$ invariance implies\Ref\EG{
         J. Ellis and M. K. Gaillard, Nucl. Phys. B150, 141 (1979).}
$$\eqalign{& A^{a,d}(P)=A^{a,d}(P)^*,\cr
           & B^{a,d}(P)=B^{a,d}(P)^*,\cr
           & C^{a,d}(P)=D^{a,d}(P)^*,\cr}\eqno(2.3)$$
where the superscripts $a$ and $d$ refer to their
absorptive and dispersive parts, respectively.
While $A^{a,d}$ and $B^{a,d}$ must be real, $C^{a,d}$ and
$D^{a,d}$ can be complex if  $CP$ is violated.
Unitarity of the  $S$-matrix  implies
$$\sum_{spin}Im\Bigl[\bar u_t(P)(-\Sigma(p))u_t(P)\Bigr]_{P\!\!\!\!/=m_t}
  =\Gamma_t,\eqno(2.4)$$
from which we have
$$A^a(m_t)+B^a(m_t)+ReC^a(m_t)={\Gamma_t\over 2m_t}.\eqno(2.5)$$
It follows from the standard on-shell renormalization prescription
that the
renormalized top-quark propagator is
$$iS_{(1)}(P)=Q_L(m_t){i\over P\!\!\!\!/-m_t+{i\over 2}\Gamma_t
     +\Sigma_{Ren}(P)}Q_R(m_t),
\eqno(2.6)$$
where
$$\eqalign{
\Sigma_{Ren}(P)&=2[A(P)-A(m_t)]P\!\!\!\!/L+2[B(P)-B(m_t)]P\!\!\!\!/R\cr
               &+m_t[C(P)-C(m_t)]L+[D(P)-D(m_t)]m_tR\cr}\eqno(2.7)$$
is the renormalized self-energy with $\Sigma_{Ren}(P)\vert_{P\!\!\!\!/=m_t}
=0$, and $Q_{L,R}$ will be defined below.
The relationship
between $m_t$ and $m_t^{(0)}$ is
$$m_t=m_t^{(0)}-m_t[A^d(m_t)+B^d(m_t)+ReC^d(m_t)],\eqno(2.8)$$
in which a regularized
 $CP$-odd phase associated with $ImC^d(m_t)\sim {\cal{O}}(\alpha)$
has been removed by a chiral rotation.  Up to this overall phase,
wave function renormalization rescales
the bare fields $\psi^0$ and $\bar\psi^0$ to
$$\eqalign{&\psi^0\to [1+A^d(m_t)L+B^d(m_t)R]\psi^0,\cr
           &\bar\psi^0\to \bar\psi^0[1+A^d(m_t)R+B^d(m_t)L].\cr}\eqno(2.9)$$
Since $A^d$ and $B^d$ are real from $CPT$ (see Eq. (2.3)),
the wave function renormalization matrix is real and hence respects $CP$.
$Q_L$ and $Q_R$ are the residuals left over by
wave function renormalization
$$\eqalign{& Q_L(m_t)=1-iA^a(m_t)L-iB^a(m_t)R+{1\over 2}ImC^a(m_t)\gamma_5,\cr
           & Q_R(m_t)=1-iA^a(m_t)R-iB^a(m_t)L+{1\over 2}ImC^a(m_t)\gamma_5,\cr}
          \eqno(2.10)$$
in which the $ImC^a(m_t)\gamma_5$ term violates $CP$.

An interesting feature emerging from this simple exercise is that
the effective lagrangian associated with (2.6) is no longer hermitian,
because $A^a(m_t), B^a(m_t), C^a(m_t)\ne 0$.
One important consequence is that the renormalized kinetic energy
part of the lagrangian does not have the standard
normalization\rlap,\Ref\HG{ For a discussion in the heavy
quark effective theory see: H. Georgi, Phys. Lett. B297, 353 (1992).}
due to $Q_{L,R}(m_t)\ne 1$.  The other is that
the relation between
the renormalized field and its hermitian adjoint differs
from that in the usual situation.
At tree level $\psi^0$ and
$\bar\psi^0$ are considered as independent variables.
Their lowest order
dynamics  are determined by the
Dirac equations,  with the solutions that $\bar\psi^0$ is
orthogonal to $\psi^0$ and that $\bar\psi^0=\psi^{0\dag}\gamma_0$.
Since wave function renormalization
does not change the degrees of freedom,
we consider that
the renormalized fields
$$\eqalign{&\psi\equiv [1+A^d(m_t)L+B^d(m_t)R]\psi^0,\cr
         &\tilde\psi\equiv \bar\psi^0[1+A^d(m_t)R+B^d(m_t)L],\cr}\eqno(2.11)$$
remain as independent fields given by their equations of motion.
For a stable fermion,
$\tilde\psi$ and $\bar\psi=\psi^{\dag}\gamma_0$
satisfy the same Dirac equation, and thus
there is no distinction between them, i.e.,
$\tilde\psi=\bar\psi$.  However, for an unstable fermion the equations
obeyed by $\tilde\psi$ and $\bar\psi$ are different (see below)
and hence
$\tilde\psi\ne\bar\psi$.  In this case, we do not
consider that the standard interpretation of wave function
renormalization $\tilde\psi=\bar\psi$ can be defined unambiguously.

The effective lagrangian associated with
(2.6) can be written in terms of
$$\eqalign{&\psi'\equiv Q_L^{-1}(m_t)\psi,\cr
           &\tilde\psi'\equiv \tilde\psi Q_R^{-1}(m_t),\cr}\eqno(2.12)$$
so that it has the standard normalization
$${\cal{L}}_{eff}^{(0)}={i\over
2}\Bigl[\tilde\psi'\gamma^{\mu}\partial_{\mu}\psi'
     -(\partial_{\mu}\tilde\psi')\gamma^{\mu}\psi'\Bigr]
     -\Bigl[m_t-{i\over 2}\Gamma_t\Bigr]\tilde\psi'\psi',\eqno(2.13)$$
where for simplicity we have not displayed the counter terms associated
with (2.9) explicitly.
We refer to $\psi'$ and $\tilde\psi'$ as `energy-eigenstate' fields,
for $\langle \tilde\psi'\vert$ and
$\vert \psi'\rangle$ are eigenstates of the Hamiltonian associated
with ${\cal{L}}_{eff}^{(0)}$.
The equations obeyed by $\psi'$ and $\tilde\psi'$
are
$$\eqalign{&i\gamma^{\mu}\partial_{\mu}\psi'(x)
-[m_t-i\Gamma_t/2)]\psi'(x)=0,\cr
&\partial_{\mu}\tilde\psi'(x)(i\gamma^{\mu})
+\tilde\psi'(x)[m_t-i\Gamma_t/2)]=0.\cr}\eqno(2.14)$$
By contrast,
$\bar\psi'$ is determined by an equation
$$\partial_{\mu}\bar\psi'(x)(i\gamma^{\mu})
+\bar \psi'(x) [m_t+i\Gamma_t/2)]=0,\eqno(2.15)$$
which is different from that obeyed by $\tilde\psi'$.
Solutions to (2.14) are summarized in Appendix A.  They
are analogous in many respects to the standard results.
With these results the physical meaning of the renormalized
propagator becomes clearer
$$\eqalign{\langle 0\vert T\psi(x)\tilde\psi(y)\vert 0\rangle&
  =\int {d^4P\over (2\pi)^4}e^{-iP\cdot (x-y)}
     iS_{(1)}(P)\cr
&\ne \langle 0\vert T\psi(x)\bar\psi(y)\vert 0\rangle,\cr}\eqno(2.16)$$
where $S_{(1)}(P)$ is given by (2.6).
The conventionally used  propagator
in the literature turns out to be
$$\langle 0\vert T\psi'(x)\tilde\psi'(y)\vert 0\rangle
=\int {d^4P\over (2\pi)^4} e^{-iP\cdot(x-y)}
 {i\over P\!\!\!\!/-m_t+i\Gamma_t/2+\Sigma_{Ren}(P)},\eqno(2.17)$$
which is neither $\langle 0\vert T\psi'(x)\bar\psi'(y)\vert 0\rangle$
nor $\langle 0\vert T\psi(x)\bar\psi(y)\vert 0\rangle$.

\chapter{Phenomenological Implications}

It has long been well known\Ref\LOY{See for example:
T. D. Lee, R. Oehme and C. N. Yang, Phys. Rev. 106, 340 (1957).}
that an unstable particle does not conjugate to its hermitian adjoint.
This feature arises automatically in the renormalization prescription
discussed above.   Instead,
$\psi'$ and $\tilde\psi'$ are orthogonal to each other (for
details see Appendix A),
and thus they provide a convenient basis for perturbation expansion
$${\cal{L}}={\cal{L}}^{(0)}_{eff}(\psi'(x), \tilde\psi'(x),
     \partial_{\mu}\psi'(x),\partial_{\mu}\tilde\psi'(x))+
    {\cal{L}}_I(\psi'(x),\tilde\psi'(x)).\eqno(3.1a)$$
The idea of using conjugate energy-eigenstate fields in
perturbation expansion is not new.  It can be traced back
to the early phenomenological studies of the neutral
kaon system
\rlap.\REFS\sachs{
          R. Jacob and R. G. Sachs, Phys. Rev. 122, 350 (1961),
          R. G. Sachs, Annals of Phys. 22, 239 (1963).}\REFSCON\EL{
          C. P. Enz and R. R. Lewis, Helv. Phys. Acta, 38, 860 (1965),
          and in {\it CP Violation}, ed. L. Wolfenstein,
          North Holland, Amsterdam (1989), p58.}
          \refsend
In the present description the state $\langle \tilde\psi'\vert$
plays exactly the same role as an `inverse state' of the
neutral kaons introduced in Refs. \sachs\ and  \EL .

In terms of $\psi'$ and $\tilde\psi'$ the free energy part of
the effective lagrangian has the standard normalization,
but  ${\cal{L}}_I$ receives a correction given by (2.12)
due to $Q_{L,R}(m_t)\ne 1$.  An alternative approach would
be to use $\psi$ and $\tilde\psi$ as the expansion basis
$${\cal{L}}^{(0)}_{eff}={\cal{L}}^{(0)}_{eff}(
\psi, \tilde\psi,\partial_{\mu}\psi,\partial_{\mu}\tilde\psi)+
\delta{\cal{L}}(A^a, B^a, ImC^a; \psi,\tilde\psi, \partial_{\mu}\psi
, \partial_{\mu}\tilde\psi).\eqno(3.1b)$$
In this picture ${\cal{L}}_I$ remains basically the same as (1.3),
and $\delta{\cal{L}}$ accounts for the changes due to (2.12).
One can show easily that these two approaches are
equivalent.  In what follows we will use (3.1a) for convenience.

Although ${\cal{L}}^{(0)}_{eff}$ is to be treated as a
`free' lagrangian, it is important to note that the perturbation
expansion given by (3.1) is different from that
in the usual situation.  The validity of this approach is briefly
discussed in Appendix B.
In many of its applications the most important step is to transform
the renormalized free energy part of the lagrangian into
a canonical form such as  (2.13).  The transformation
rules are given in (2.12).  Corrections
arising from kinematics and from  modifications to the Dirac
algebra, which are  discussed in Appendix A,  often turn out to be
unimportant in calculating $CP$ asymmetries.

For the case discussed in Introduction we find that
wave function renormalization introduces an order $\alpha$
correction to the decay amplitudes.  The results are
$$\eqalign{{\cal{M}}(t\to bW^+)&={ig\over \sqrt 2}\epsilon^{\mu}
\bar u_b\gamma_{\mu}LQ_L(m_t)u'_t\cr
&={ig\over\sqrt 2}\Bigl[1-iA^a(m_t)-{1\over 2}ImC^a(m_t)\Bigr]
  \epsilon^{\mu}\bar u_b\gamma_{\mu}Lu'_t,\cr}\eqno(3.2)$$
where $\epsilon^{\mu}$ is the $W$ polarization, and
$$\eqalign{{\cal{M}}(\bar t\to \bar b W^-)={ig\over\sqrt 2}&
\Bigl[1-iA^a(m_t)+{1\over 2}ImC^a(m_t)\Bigr]\epsilon^{\mu}
\tilde v'_t\gamma_{\mu}Lv_b,\cr
{\cal{M}}(t\to bH^+)={ig\over\sqrt 2}&\bar u_b\Bigl[\Bigl[
   1-iA^a(m_t)-{1\over 2}ImC^a(m_t)\Bigr]xm_bL\cr
 &+
   \Bigl[1-iB^a(m_t)+{1\over 2}ImC^a(m_t)\Bigr]ym_tR\Bigr]u'_t,\cr
{\cal{M}}(\bar t\to\bar bH^-)={ig\over\sqrt 2}&\tilde v'_t\Bigl[\Bigl[
 1-iA^a(m_t)+{1\over 2}ImC^a(m_t)\Bigr]x^*m_bR\cr
 &+
 \Bigl[1-iB^a(m_t)-{1\over 2}ImC^a(m_t)\Bigr]y^*m_tL\Bigr]v_b.\cr}\eqno
(3.3)$$
 In obtaining these results we have assumed for simplicity that
$b$ and $W$ are stable.
Diagrammatically these corrections correspond to contributions generated from
one-particle reducible graphs connecting  an external self-energy
to a vertex $t\to bW, bH$ (Figs. 1e and 1f).

In the charged-Higgs-boson decay,
the interference between the $CP$-violating phase $Im(x^*y)$
and the $CP$-conserving phases $iA^a(m_t)$ and $iB^a(m_t)$ in (3.3) produces
a $CP$ asymmetry.  To lowest order in $x$ and $y$,
the relevant contribution to $A^a$ and $B^a$ comes from
an intermediate $W$ in the $t$-quark self-energy with
$$\eqalign{&A^a(m_t)={\Gamma(t\to bW)\over 2m_t},\cr
&B^a(m_t)=0.\cr}\eqno(3.4)$$
Substituting (3.4) into (3.3) we find
$$\Delta_H(wave\ function)=-{g^2m_b^2Im(x^*y)\Gamma(t\to bW)\over
16\pi M_W^2m_t^2}\sqrt{\lambda(m_t^2, M_H^2, m_b^2)}.\eqno(3.5)$$
Effectively, $\Delta_H(wave\ function)$ is generated from an interaction
in which $t$ decays to an intermediate $bW^+$ state followed by a
$CP$-violating final-state $s$-channel scattering $bW^+\to bH^+$
(Fig. 1e).

In the $W$-boson decay channel the $CP$-violating term
$ImC^a(m_t)$ induces a $CP$ asymmetry by itself
$$\Delta_W(wave\ function)=-2ImC^a(m_t)\Gamma(t\to bW).\eqno(3.6)$$
In this case the relevant interaction is induced by a
decay $t\to bH^+$ followed by an
$s$-channel final-state scattering $bH^+\to bW^+$ (Fig. 1f).
In the model considered in this paper a simple calculation shows
$$ImC^a(m_t)=-{g^2m_b^2Im(x^*y)\over 32\pi M_W^2m_t^2}
\sqrt{\lambda(m_t^2, M_H^2, m_b^2)}.\eqno(3.7)$$
Comparing Eqs. (3.5), (3.6) and (3.7) one sees that
$$\Delta_W(wave\ function)=-\Delta_H(wave\ function).\eqno(3.8)$$
It follows from (1.5) and (3.8) that
$CPT$ invariance is restored.
In the present description mass renormalization
does not contribute to the $CP$ asymmetry, in contrast to (1.7).

Without applying (1.6) one will not encounter the
$CPT$ violation difficulty.  However, a naive application
of the standard calculation is still incomplete.  Indeed,
if we were to follow the standard interpretation
of wave function renormalization we would have identified
$\tilde\psi$ as $\bar\psi$.  In that case, a linear transformation
$\psi^{(0)}\to Q_L\psi$ would have implied  $\bar\psi^{(0)}
\to \bar\psi\gamma_0Q_L^{\dag}
\gamma_0$, and as a consequence we would have
concluded after a
simple calculation that $\Delta_W(wave\ function)=\Delta_H(wave\ function)
=0$ (Eq. 1.8).
This is probably the reason why wave-function renormalization
effects on $CP$-violating asymmetries have hardly been discussed
before.

Potentially the missing piece (3.8)  can be of interest:
due to the non-universality of the charged-Higgs-boson interaction,
a large $\Delta_W=\Delta_W(vertex)
+\Delta_W(wave\ function)=-\Delta_H$ may induce a large $CP$ asymmetry
in the top-quark semileptonic decays without
a small lepton mass suppression.  The effect due to
wave function renormalization alone is
$$\sum_{\ell=e,\mu}
{\Gamma(t\to b\bar\ell\nu_{\ell})-\Gamma(\bar t\to \bar b\ell\bar\nu_{\ell})
\over \Gamma(t\to bW^+)+\Gamma(\bar t\to \bar bW^-)}
\approx 3\times 10^{-5}Im(x^*y).\eqno(3.9)$$
The complete result for this asymmetry
parameter depends on other details of the model not specified by (1.3).

Phenomenologically, the $CP$-violating term $ImC^a(m_t)\gamma_5$
is originated from $CP$ violation in the decay matrix.
Hence interactions associated with this term automatically have
both $CP$-violating and $CP$-conserving phases.
Besides the effect discussed above,
the $ImC^a(m_t)\gamma_5$ term can generate $CP$-violating
observables in processes involving the    operators
listed below.  Neglecting the $CP$-conserving phases $A^a(m_t)$
and $B^a(m_t)$, (2.12) is simplified to
$$\eqalign{&\psi^{(0)}\to \Bigl[1+{1\over 2}ImC^a(m_t)\gamma_5\Bigr]\psi',\cr
     &\bar\psi^{(0)}\to\tilde\psi'\Bigl[1+{1\over 2}ImC^a(m_t)
\gamma_5\Bigr].\cr
}\eqno(3.10)$$
For an asymmetric fermion pair we have
$$\eqalign{&\bar\psi_f^{(0)}\gamma_{\mu}L\psi^{(0)}\to
\Bigl[1-{1\over 2}ImC^a(m_t)\Bigr]\bar\psi'_f\gamma_{\mu}L\psi',\cr
&\bar\psi^{(0)}\gamma_{\mu}L\psi_f^{(0)}\to
\Bigl[1+{1\over 2}ImC^a(m_t)\Bigr]\tilde\psi'\gamma_{\mu}L\psi'_f,\cr}
\eqno(3.11)$$
where $f$ represents a stable fermion, i.e., $f\ne t$.
Also,
$$\eqalign{&\bar\psi_f^{(0)}\gamma_{\mu}R\psi^{(0)}\to
\Bigl[1+{1\over 2}ImC^a(m_t)\Bigr]\bar\psi_f'\gamma_{\mu}R\psi',\cr
&\bar\psi^{(0)}\gamma_{\mu}R\psi_f^{(0)}\to
\Bigl[1-{1\over 2}ImC^a(m_t)\Bigr]\tilde\psi'\gamma_{\mu}R\psi'_f;\cr
&\bar\psi^{(0)}_f\sigma_{\mu\nu}L\psi^{(0)}\to
\Bigl[1-{1\over 2}ImC^a(m_t)\Bigr]\bar\psi'_f\sigma_{\mu\nu}L\psi',\cr
&\bar\psi^{(0)}\sigma_{\mu\nu}R\psi^{(0)}_f\to
\Bigl[1+{1\over 2}ImC^a(m_t)\Bigr]\tilde\psi'\sigma_{\mu\nu}R\psi'_f.\cr}
\eqno(3.12)$$
For a symmetric fermion pair the results are
$$\eqalign{&\bar\psi^{(0)}L\psi^{(0)}\to
\Bigl[1-ImC^a(m_t)\Bigr]\tilde\psi'L\psi',\cr
&\bar\psi^{(0)}R\psi^{(0)}\to
\Bigl[1+ImC^a(m_t)\Bigr]\tilde\psi'R\psi';\cr
& \bar\psi^{(0)}\sigma_{\mu\nu}L\psi^{(0)}\to
\Bigl[1-ImC^a(m_t)\Bigr]\tilde\psi'\sigma_{\mu\nu}L\psi',\cr
&\bar\psi^{(0)}\sigma_{\mu\nu}R\psi^{(0)}\to
\Bigl[1+ImC^a(m_t)\Bigr]\tilde\psi'\sigma_{\mu\nu}R\psi'.\cr}
\eqno(3.13)$$
Interactions involving these operators can generate
a $CP$-violating asymmetry in the partial decay rate difference
or in the final-state spectrum  via spin-correlation or
distribution.  They may have interesting phenomenological implications
in searching for $CP$ violation observables\rlap,
\REFS\Europe{
            F. Hoogeveen and L. Stodolsky, Phys. Lett. B212, 505 (1988);
            W. Bernreuther and O. Nachtmann, Phys. Rev. Lett. 63, 2787 (1989);
            F. Hoogeveen, Nucl. Phys. B341, 322 (1990);
            A. Pilaftsis and M. Nowakowski, Phys. Lett.  B245, 185 (1990);
            \splitout
            W. Bernreuther, O. Nachtmann, P. Overmann and T. Schroder,
            Heidelberg preprint, HD-THEP-92-14 (1992);
            A. Pilaftsis and M. Nowakowski, preprint, MZ-TH/92-56 (1992);
            J. L. Diaz-Cruz and G. Lopez Castro, Phys. Lett. B301, 405 (1993).}
\REFSCON\USA{
            G. Eilam, J. L. Hewett and A. Soni, Phys. Rev. Lett. 67,
            1970 (1991), {\it{ibid}} 68, 2103 (1992);
            A. Aeppli, D. Atwood and A. Soni, BNL-47355, (1992);
            D. Chang and W. Y. Keung, FERMILAB-PUB-92/172-T (1992);
            D. Chang, W. Y. Keung and I. Phillips, CERN-TH-6658/92, (1992);
            R. Cruz, B. Grzadkowski and J. F. Gunion, Phys. Lett. B289,
                                       440 (1992);
            N. G. Deshpande et. al., Phys. Rev. D43, 3591 (1991);
            N. G. Deshpande, B. Morgolis and H. D. Trottier, Phys. Rev. D45,
                            178 (1992);
  \splitout
            G. Grzadkowski and J. F. Gunion, Phys. Lett. B287, 237 (1992);
            {\it ibid} B294, 361 (1992);
            C. J.-C. Im, G. L. Kane and P. J. Malde,
            University of Michigan preprint, UM-TH-92-27, (1992);
            G. L. Kane, G. A. Landinsky and C. P. Yuan, Phys. Rev. D45,
                                       124 (1992);
           C. R. Schmidt and M. E. Peskin, Phys. Lett. 69, 410 (1992);
           C. R. Schmidt, Phys. Lett. B293, 111 (1992);
           A. Soni and R. M. Xu, Phys. Rev. Lett. 69, 33 (1992).}
           \refsend
 should $ImC^a$ turn out
to be sufficiently large.  In the model discussed in this paper $ImC^a(m_t)$
is suppressed by $m_b$ (see (3.7)).
As far as their magnitudes are concerned,
it is much easier for $ImC^a$ to be
competitive in interactions involving an
asymmetric fermion pair than in that involving a symmetry pair.
A quantitative study should be done carefully.
A similar general discussion
can be given to  the $CP$-conserving phases. However, by themselves
they do not violate $CP$.

\chapter{Conclusion}

We have investigated effects introduced by the instability
of an unstable fermion on wave function renormalization,
and arrived at a phenomenological description.
In many of its applications, the most
important step is to transform the renormalized free energy part
of the lagrangian into a canonical from such as that given by (2.13).
This prescription provides a practical way of computing
one-particle reducible graphs associated with an external
self-energy bubble in the on-shell renormalization scheme.
Applications are given to the $CP$-violating asymmetries in the
$t$ decays, in which we have shown that a naive
application of the standard calculation either is incomplete or
violates $CPT$.  It is emphasized that wave function renormalization
can introduce important $CP$-conserving and $CP$-violating phases.
The present description is limited by the requirement that
the width of the unstable particle in question is small.

An interesting feature shown by this analysis is that a renormalized
unstable fermion does not conjugate to its hermitian adjoint. Therefore,
in the study of $CP$ violation it is important to distinguish
$\tilde\psi'$ from $\bar\psi'$.  We point out that this
feature does not pertain only to fermions; it happens to all particles
whenever effects arising from particle-width are not negligible.
A detailed discussion on the
phenomenological  description of an unstable boson
will be presented elsewhere.

\ack

I wish to thank P. Langacker, G. Segre and L. Wolfenstein for
many invaluable discussions.  This work was supported in part
by an $SSC$ fellowship from Texas National Research Laboratory Commission.
\endpage

\centerline{APPENDIX A}

The expansion into creation and annihilation operators of
the  solutions to (2.14) can be written as
$$\eqalign{&
\psi'(x)=\int{d^3k\over(2\pi)^3}\Bigl({m_t-i\Gamma_t/2\over k_0}\Bigr)
\sum_{\beta}\Bigl[b_{\beta}(k)u'_{\beta}(k)e^{-ikx}
+\tilde d_{\beta}(k)v'_{\beta}(k)e^{ikx}\Bigr],\cr
&\tilde\psi'(x)=\int{d^3k\over(2\pi)^3}\Bigl({m_t-i\Gamma_t/2\over k_0}\Bigr)
\sum_{\beta}\Bigl[\tilde b_{\beta}(k)\tilde u'_{\beta}(k)e^{ikx}
+d_{\beta}(k)\tilde v'_{\beta}(k)e^{-ikx}\Bigr],\cr}\eqno(A.1)$$
where $k_0$ is complex (in the rest frame of the  $t$, $k_0=m_t-i\Gamma/2$).
In the Dirac representation  the spinors are
$$\eqalign{&
u'_{\beta}(k)={1\over\sqrt{(k_0+m_t-i\Gamma_t/2)^2-\vec{k}^2}}\pmatrix{
     (k_0+m_t-i\Gamma_t/2)\phi_{\beta}\cr
      \vec{\sigma}\cdot\vec{k}\phi_{\beta}\cr},\cr
&v'_{\beta}={1\over\sqrt{(k_0+m_t-i\Gamma_t/2)^2-\vec{k}^2}}\pmatrix{
      \vec\sigma\cdot\vec{k}\phi_{\beta}\cr
     (k_0+m_t-i\Gamma_t/2)\phi_{\beta}\cr},\cr}\eqno(A.2)$$
where $\phi_{\beta}$ is the standard two-component spinor.
$\tilde u'_{\beta}$ and $\tilde v'_{\beta}$ are related to
$u'_{\beta}$ and $v'_{\beta}$ by
$$\eqalign{&\tilde u'_{\mp}(k)=\pm[Cv'_{\pm}(k)]^T,\cr
&           \tilde v'_{\mp}(k)=\mp[Cu'_{\pm}(k)]^T,\cr}\eqno(A.3)$$
where $C=i\gamma_2\gamma_0$ is the standard charge conjugation matrix.
A remarkable feature of these solutions is that
$\tilde\psi'$  and  $\psi'$ conjugate to each other
$$\eqalign{&\tilde u'_{\beta}(k)u'_{\lambda}(k)=\delta_{\beta\lambda},\cr
&\tilde v'_{\beta}(k)v'_{\lambda}(k)=-\delta_{\beta\lambda},\cr
&\tilde u'_{\beta}(k)v_{\lambda}(k)=\tilde v'_{\beta}(k)u'_{\lambda}(k)=0.
\cr}\eqno(A.4)$$
As a result, one can quantize (A.1) by introducing
anticommutators
$$\eqalign{&\{b_{\beta}(k), \tilde b_{\lambda}(k')\}
  =(2\pi)^3{k_0\over m_t-i\Gamma_t/2}\delta^3(\vec k-\vec
k')\delta_{\beta\lambda},\cr
&\{d_{\beta}(k), \tilde d_{\lambda}(k')\}
=(2\pi)^3{k_0\over m_t-i\Gamma_t/2}\delta^3(\vec k-\vec
k')\delta_{\beta\lambda},
\cr}\eqno(A.5)$$
and the others are zero.
Since $\psi'$ does not conjugate to $\bar\psi'$, i.e.,
$\bar u'_{\alpha}(k)v'_{\beta}(k)\ne \delta_{\alpha\beta}$,
it is important to note that $\{b^{\dag}_{\alpha},\tilde d_{\beta}\}\ne 0$.

The projection operators
$$\eqalign{&\Lambda_+(k)=\sum_{\alpha=\pm}u'_{\alpha}(k)\otimes
\tilde u'_{\alpha}(k)={k\!\!\!/+m_t-i\Gamma_t/2\over
    2(m_t-i\Gamma_t/2)},\cr
& \Lambda_-(k)=-\sum_{\alpha=\pm}v'_{\alpha}(k)\otimes\tilde v'_{\alpha}(k)
={-k\!\!\!/+m_t-i\Gamma_t/2\over 2(m_t-i\Gamma_t/2)},\cr}\eqno(A.6)$$
have the standard properties
$$\eqalign{&\Lambda_+(k)+\Lambda_-(k)=1,\cr
&Tr[\Lambda_{\pm}(k)]=2,\cr
&\Lambda_{\pm}^2(k)=\Lambda_{\pm}(k),\cr
&\Lambda_+(k)\Lambda_-(k)=\Lambda_-(k)\Lambda_+(k)=0.\cr}\eqno(A.7)$$
The projection operators which enter into
the calculation of the decay rates studied in the context are
$$\eqalign{&\sum_{\alpha=\pm}u'_{\alpha}(k)\otimes \bar u'_{\alpha}(k)
={k\!\!\!/+m_t-i\Gamma_t/2\over\vert (k_0+m_t-i\Gamma_t/2)^2
-\vec k^2\vert}\pmatrix{k_0^*+m_t+i\Gamma_t/2 & 0\cr
                        0 & k_0+m_t-i\Gamma_t/2\cr},\cr
&\sum_{\alpha=\pm}\gamma_0\tilde v_{\alpha}^{\prime \dag}(k)
\otimes\tilde v'_{\alpha}(k)=\pmatrix{k_0+m_t-i\Gamma_t/2 & 0\cr
                             0 & k_0^*+m_t+i\Gamma_t/2\cr}
{k\!\!\!/-m_t+i\Gamma_t/2\over
\vert (k_0+m_t-i\Gamma_t/2)^2-\vec k^2\vert},\cr}\eqno(A.8)$$
in which each entry represents a $2\times 2$ block.  All the results
presented above reduce to the standard forms in the limit $\Gamma_t=0$.
\endpage
\centerline{APPENDIX B}

Justifications for perturbation expansions like
those given by (3.1a) and (3.1b) have been discussed extensively
in Refs. \sachs\ and \EL .  Here we simply extend their arguments
to fermions in field theory.

It is important to note that the energy-eigenstate fields $\psi'$
and $\tilde\psi'$ are not physical, i.e., they cannot be prepared
or observed.  This follows because $\psi'$ and $\tilde\psi'$ do
not conjugate to their hermitian adjoints.  As a result, it is
impossible to interpret $\bar\psi'\gamma_0\psi'$ and
$\tilde\psi'\tilde\psi^{\prime\dag}$ as particle density operators.
In fact, since ${\cal{L}}_{eff}^{(0)}$ is invariant under
the transformation
$$\eqalign{&\psi'\to \lambda\psi',\cr
           &\tilde\psi'\to \lambda^{-1}\tilde\psi',\cr}\eqno(B.1)$$
where $\lambda$ is arbitrary, the orthonormality condition
between $\psi'$ and $\tilde\psi'$ leaves undetermined not only
their phases but also their scales.  In Appendix A we have
chosen $\lambda=1$ for convenience.

As pointed out in Ref. \sachs\ and particularly in Ref. \EL ,
the resolution of these difficulties lies in the recognition that
the energy-eigenstate fields $\psi'$ and $\tilde\psi'$ have
only an intermediate role in calculation.  They appear only in
the expansion of physical states via
$$\vert \tilde\psi'\rangle\langle \psi'\vert=1.\eqno(B.2)$$
Thus, the ambiguity in the normalization of $\psi'$ and $\tilde\psi'$
does not enter in the determination of physical amplitudes.

For example, in the decay $t\to F$, where $F$ is either $bW$ or
$bH$, the amplitude for detecting a $F$
at a time $\tau_f$ from the decay of $t$ produced at an earlier time $\tau_i$
is
$$\eqalign{{\cal{M}}(t\to F)&=\langle F;\tau_f\vert t;\tau_i\rangle\cr
&=\langle F;\tau_f\vert
\tilde\psi';\tau_i\rangle\langle \psi';\tau_i\vert t
; \tau_i\rangle\cr
&=\langle F;\tau_f\vert e^{-iH_I(\tau_f-\tau_i)}\vert\tilde\psi';\tau_f
\rangle e^{-iP_t^0(\tau_f-\tau_i)}\langle \psi';\tau_i\vert
t;\tau_i\rangle,\cr}\eqno(B.3)$$
where
$$H_I=-\int d^3x{\cal{L}}_I(\psi'(x),\tilde\psi'(x)),\eqno(B.4)$$
and $P_t^{0}=m_t-i\Gamma_t/2$ in the rest frame of $t$.
In practice,
the extra factor $\langle\psi';\tau_i\vert t;\tau_i\rangle$
usually introduces a trivial effect.
It cancels in the ratio $\Delta_{W,H}/\Gamma(t\to all)$.

The remaining factor in (B.3) is  related to a Green's function
by a relation similar to the standard reduction formula.
For $F=b(P_b;\alpha)W(P_W)$ the final result is
$$\eqalign{&\langle b(P_b,\alpha)W(P_W);\tau_f\vert
e^{-iH_I(\tau_f-\tau_i)}\vert
 \tilde\psi'(P_t,\beta);
\tau_f\rangle\cr
&\approx (2\pi)^3\delta^3
(\vec P_t-\vec P_b-\vec P_W)\theta(\tau_f-\tau_i)
\int_{\tau_i}^{\tau_f}d\tau e^{-i\tau(P_t^0-P_b^0-P_W^0)}
{ig\over \sqrt 2}
\bar u_{b,\alpha}\gamma_{\mu}LQ_Lu'_{t,\beta}
\epsilon^{\mu}\cr
&\approx (2\pi)^4\delta^3(\vec P_t-\vec P_b-\vec P_W)\delta(
ReP_t^0-P_b^0-P_W^0){ig\over \sqrt 2}\bar u_{b,\alpha}
\gamma_{\mu}LQu'_{t,\beta}
\epsilon^{\mu}.\cr}\eqno(B.4)$$
In the last step of (B.4) as well as in obtaining the $\theta$-function
we have explicitly assumed that $\Gamma_t\ll m_t $.  This
is the condition for the validity of the results presented in this
paper.
\endpage
\centerline{FIGURE CAPTION}

Fig.1.  Feynman diagrams for the $CP$-violating asymmetries
in the $t$ decays, in which  $\phi^0$ represents a neutral boson.

\endpage

\refout
\end